\documentclass[%
 reprint,
 amsmath,amssymb,
 aps,
pra,
]{revtex4-1}

\usepackage{float}
\usepackage[usenames, dvipsnames]{color}
\usepackage{graphicx}% Include figure files
\usepackage{dcolumn}% Align table columns on decimal point
\usepackage{bm}% bold math
\begin{document}

\preprint{APS/123-QED}

\title{Revisiting photon-statistics effects on multiphoton ionization. II.}% Force line breaks with \\

\author{G. Mouloudakis$^1$}
 \email{gmouloudakis@physics.uoc.gr}

\author{P. Lambropoulos$^{1,2}$}%

\affiliation{${^1}$Department of Physics, University of Crete, P.O. Box 2208, GR-71003 Heraklion, Crete, Greece
\\
${^2}$Institute of Electronic Structure and Laser, FORTH, P.O.Box 1527, GR-71110 Heraklion, Greece}

\date{\today}% It is always \today, today,

\begin{abstract}
In this paper, we extend the results of an earlier paper in which we had demonstrated the limitations of the notion of nonresonant multiphoton ionization, in the exploration of photon statistics effects in non-linear processes. Through the quantitative analysis of specific realistic processes, we provide the connection to conditions of intensity and pulse duration necessary in relevant experiments, including a recent seminal experiment demonstrating the effect of superbunching found in squeezed radiation.

\begin{description}

\item[PACS numbers]
32.80.Rm, 42.50.Ar

\end{description}
\end{abstract}

\maketitle

%\tableofcontents

\section{\label{sec:level1}Introduction}

In a recent paper \cite{ref1}, referred to hereafter as I, we examined the effect of field fluctuations on near-resonant 2- and 3-photon ionization. Through a quantitative analysis of the role of intensity, we found that the enhancement of the process due to bunching, more often than not, will be reduced owing to the onset of Rabi oscillations, even at seemingly large detuning from resonance. This makes the notion of non-resonant ionization rather tenuous and possibly misleading in the planning of an experiment. For processes of order higher than 3, it is practically inevitable that near resonances will modify the enhancement, even at moderate intensities; simply because the level spacing decreases with increasing level excitation.

That paper came at an auspicious time as it practically coincided with an experimental breakthrough by Spasibko  et al. \cite{ref2}, in which enhancement in processes of order up to 4 due to superbunched light was reported. This represents a break in the rather long period of experimental drought in this field. Because, despite the    
rather extensive relevant theoretical literature over the
 last 50 years \cite{ref3,ref4,ref5,ref6,ref7,ref8,ref9,ref10}, experimental results have been quite sparse \cite{ref11,ref12,ref13,ref14,ref15,ref16,ref17}. But most significantly because photon statistics enhancement due to squeezed (superbunched) light has been observed for the first time; a development that opens the route towards the experimental investigation of a number of challenging open problems.

Owing to computational demands, the numerical results in I, illustrating the involvement of near-resonances and the ensuing modification of the bunching effect with increasing intensity, were obtained with scaled intensity and atomic parameters; such as Rabi frequencies and ionization cross sections. The need for scaling arose from the necessity of formulating the problem in terms of a quantized field, which then led to summations over the photon number distributions of the various states of the field we examined. The concomitant drawback of that approach is the difficulty in translating photon numbers to intensity of a pulsed source, which is what enters an experiment \cite{ref2}. Our purpose in this article is to provide a bridge between the two, through a sample of realistic calculations in atomic systems. 

For that purpose, we have chosen the cases of 2-photon ionization in Cesium in the vicinity of the intermediate 7p state and 3-photon ionization in Sodium, in the vicinity of 2-photon intermediate resonance with the 4d(3/2) and 4d(5/2) states. It is common knowledge that, for a non-resonant N-photon process, the ionization yield as a function of laser intensity, in a log-log plot, will be a straight line of slope N \cite{ref3}. When, with increasing intensity, an intermediate state near resonance begins playing a role, the slope begins diminishing. The intensity at which this change of slope becomes noticeable is the intensity at which the bunching enhancement will also begin decreasing. 

The most convenient formal tool for the calculation of the  dependence of ion yields on intensity  is the time-dependent density matrix of the atomic system, driven by a classical electromagnetic (EM) field. This formulation, in addition to allowing for realistic calculations which include the source temporal pulse shape, it does also allow for the incorporation of the laser bandwidth; an experimentally important aspect not accessible in the formalism of I. Evaluating the role of the bandwidth addresses the possibility that, even with the carrier frequency tuned far from resonance, the wings of the spectral shape of the pulse may involve real excitation of a near resonant state, thereby altering the slope and therefore the enhancement due to bunching.

A short clarification of the issue of bandwidth is needed at this point. The bandwidth of a pulsed source involves two separate contributions. One is due to the finite duration of the pulse and is referred to as the Fourier bandwidth. The second is due to stochastic phase and/or amplitude fluctuations of the field; the latter being the cause of bunching. Depending on the duration of the pulse, the stochastic bandwidth may become important and therefore would have to be included in the formulation. For the sake of completeness we have included the effects of the stochastic bandwidth in our calculations.

For pulses of extremely short duration of the order of a few femtoseconds, it is the Fourier bandwidth that dominates. For longer pulses of hundreds of femtoseconds or longer, it is the stochastic bandwidth that dominates, which is in fact the case with sources in which bunching is significant \cite{ref2}. In the examples discussed in this paper, the pulses we have chosen are of relatively long duration, so that it is the stochastic bandwidth that dominates. Nevertheless, the Fourier bandwidth is inherently included in the calculation, through the solution of the time-dependent density matrix differential equations.

In order to avoid repetition of formal aspects and derivations readily available in the literature \cite{ref18}, we have chosen to simply provide the basic differential equations governing the time evolution of the density matrix for the 3-photon process, as a special case of which the 2-photon process can be obtained. After a brief discussion and explanation of the basic equations in section II, a collection of illustrative examples with discussion and conclusions are presented in section III.

\section{\label{sec:level2}Theory}

Consider the atom to be initially in its ground state, denoted by $\left| 1 \right\rangle$, in the presence of an external electric field of the form $E(t) = \mathcal{E} (t){e^{i\omega t}} + c.c.$, where $\omega$ is the frequency of the field and $\mathcal{E} (t)$ its complex amplitude. The field amplitude can be expressed as $\mathcal{E} (t) = \left| {\mathcal{E} (t)} \right|\exp [i\varphi (t)]$, where both $\left| {\mathcal{E} (t)} \right|$ and $\varphi (t)$ can in general be stochastic quantities, owing to fluctuations of the field. The absorption of two photons  excites the atom to the vicinity of an intermediate state $\left| 2 \right\rangle$, whose energy is denoted by ${\omega _2} \left( {\hbar  = 1} \right)$. In addition to the coherent coupling between the ground and excited state induced by the external field, the state $\left| 2 \right\rangle $ can either decay spontaneously or be ionized by a single photon absorption, with a rate denoted by $\Gamma(t)$. When state $\left|2\right\rangle$ is connected to the initial state by a two-photon transition, spontaneous decay back to $\left|1\right\rangle$ is allowed only through a cascade. For the sake of completeness we include  in our equations that rate, in the sense of an effective rate denoted by $\gamma$. In any case, for the type of experiments pertaining to our problem, that rate is too small to be of relevance. The excited state $\left|2\right\rangle$, in addition to its coupling to the initial state and to the continuum, it does also undergo a Stark shift through virtual transitions to all dipole allowed bound and continuum states. It is denoted here by $S(t)$, where the time dependence is due to the fact that the shift is proportional to  the instantaneous intensity. Since in this case, the two-photon Rabi frequency, as well as the ionization rate, are also proportional to the intensity, the shift needs to be included in the formalism.

If ${\rho _{ij}}$, $i,j=1,2$ are the matrix elements of the density matrix of our effective two-level model of the three-photon process, in the rotating wave approximation (RWA), then the differential equations governing the time evolution of the respective slowly-varying matrix elements, defined by ${\rho _{ii}}(t) = {\sigma _{ii}}(t)$, $i=1,2$ and ${\rho _{12}}(t) = {\sigma _{12}}(t){e^{i\omega t}}$ are \cite{ref1}:
\begin{equation}
{\partial  \over {\partial t}}{\sigma _{11}}\left( t \right) = \gamma {\sigma _{22}}\left( t \right) + 2{\mathop{\rm Im}\nolimits} \left[ {\Omega _{12}^*\left( t \right){\sigma _{12}}\left( t \right)} \right]
\end{equation}
\begin{equation}
{\partial  \over {\partial t}}{\sigma _{22}}\left( t \right) =  - \left[ {\gamma  + \Gamma \left( t \right)} \right]{\sigma _{22}}\left( t \right) - 2{\mathop{\rm Im}\nolimits} \left[ {\Omega _{12}^*\left( t \right){\sigma _{12}}\left( t \right)} \right]
\end{equation}
\begin{multline}
{\partial  \over {\partial t}}{\sigma _{12}}\left( t \right) = \left\{ {i\left[ {\Delta  - S(t)} \right] - {\gamma _{12}}\left( t \right)} \right\}{\sigma _{12}}\left( t \right) \\ + i{\Omega _{12}}\left( t \right)\left[ {{\sigma _{22}}\left( t \right) - {\sigma _{11}}\left( t \right)} \right]
\end{multline}
where $\Delta  = 2\omega  - ({\omega _2} - {\omega _1})$ is the detuning from the intermediate resonance, ${\gamma _{12}}\left( t \right) = {1 \over 2}\left[ {\gamma + \Gamma \left( t \right)} \right]$ is the off-diagonal relaxation and ${\Omega _{12}}\left( t \right) = {\hbar ^{ - 2}}{\mu _{12}}\mathcal{E}^{2}(t)$ is the effective two-photon Rabi frequency of the $\left| 1 \right\rangle  \leftrightarrow \left| 2 \right\rangle $ transition, given by the product of the effective two-photon matrix element ${{\mu _{12}}}$ of the dipole operator and the square of the electric field amplitude.

All parameters depending on the applied EM field, namely the ionization rate, the Stark shift and the Rabi frequency, are in general stochastic quantities, as they are subjected to the fluctuations of the field. As a result, the equations of motion of the density operator's matrix elements are also stochastic. The next step is to average the differential equations (1) to (3) over the field fluctuations, denoting such averaged quantities by angular  brackets. In the process of averaging the above set of differential equations, we encounter atom-field products of the form $\left\langle {\Gamma (t){\sigma _{ij}}\left( t \right)} \right\rangle $, $\left\langle {{\Omega _{12}}(t){\sigma _{ij}}\left( t \right)} \right\rangle $, etc., $i,j=1,2$. The rigorous evaluation of such products require the exact model of the stochastic properties of the field, entailing considerable mathematical complexity, which is beyond the scope of this paper. However, a detailed treatment of how this can be accomplished can be found in reference \cite{ref5}.  Since our purpose in this paper is to simply assess the effect of the laser bandwidth, in an approximate fashion, we adopt the decorrelation approximation, which amounts to replacing the stochastic averages of products such as the above, by the products of their averages. Assuming that the deterministic real field amplitude is constant and equal to $\mathcal{E}_0$ (square pulse shape), the resulting differential equations governing the time-evolution of the averaged density matrix elements are:

\begin{widetext}
\begin{equation}
{\partial  \over {\partial t}}\left\langle {{\sigma _{11}}(t)} \right\rangle  = \gamma \left\langle {{\sigma _{22}}(t)} \right\rangle  + 2{\mathop{\rm Im}\nolimits} \left\{ {i\bar \Omega _{12}^2\int\limits_0^t {dt'{e^{i\left[ {\Delta  - \left\langle S \right\rangle  - \left\langle {{{\tilde \gamma }_{12}}} \right\rangle } \right]\left( {t - t'} \right)}}\left[ {\left\langle {{\sigma _{22}}(t')} \right\rangle  - \left\langle {{\sigma _{11}}(t')} \right\rangle } \right]} } \right\}
\end{equation}
\begin{equation}
{\partial  \over {\partial t}}\left\langle {{\sigma _{22}}(t)} \right\rangle  =  - \left[ {\gamma  + \left\langle \Gamma  \right\rangle } \right]\left\langle {{\sigma _{22}}(t)} \right\rangle  - 2{\mathop{\rm Im}\nolimits} \left\{ {i\bar \Omega _{12}^2\int\limits_0^t {dt'{e^{i\left[ {\Delta  - \left\langle S \right\rangle  - \left\langle {{{\tilde \gamma }_{12}}} \right\rangle } \right]\left( {t - t'} \right)}}\left[ {\left\langle {{\sigma _{22}}(t')} \right\rangle  - \left\langle {{\sigma _{11}}(t')} \right\rangle } \right]} } \right\}
\end{equation}

\end{widetext} 
where $\left\langle {{{\tilde \gamma }_{12}}} \right\rangle  \equiv \left\langle {{\gamma _{12}}} \right\rangle  + {\gamma _L} = {1 \over 2}\left( {\gamma  + \left\langle \Gamma  \right\rangle  + 2{\gamma _L}} \right)$ and ${{\bar \Omega }_{12}} \equiv {\hbar ^{ - 1}}{\mu _{12}}{\mathcal{E}_0^2}$ the average 
Rabi frequency of the $\left| 1 \right\rangle  \leftrightarrow \left| 2 \right\rangle $ transition.

Note that $\left\langle {\Gamma (t)} \right\rangle $ and $\left\langle {S(t)} \right\rangle $ have been replaced by $\left\langle {\Gamma} \right\rangle $ and $\left\langle {S} \right\rangle $, respectively, since we assumed that the deterministic field amplitude is constant.
The above averaged equations contain the additional term ${{\gamma_L}}$ in the off-diagonal relaxation which represents the bandwidth of the radiation source. This result is the consequence of relations \cite{ref19,ref20,ref21} that connect higher-order field correlation functions, to the second-order field correlation function:
\begin{equation}
\left\langle {\mathcal{E}\left( {{t_1}} \right){\mathcal{E}^*}\left( {{t_2}} \right)} \right\rangle  = \left\langle {{\mathcal{E}^2}} \right\rangle\exp \left[ { - {1 \over 2}{\gamma _L}\left| {{t_1} - {t_2}} \right|} \right]
\end{equation}
where $ \left\langle {{\mathcal{E}^2}} \right\rangle $ is the variance of the electric field.
Such field correlation functions appear in the process of averaging over field's stochastic fluctuations and introduce the laser bandwidth in our model.

The solution of the equations of motion of the averaged density matrix elements, provide the probability of ionization through the expression

\begin{equation}
P(t) = 1 - \left\langle {{\sigma _{11}}(t)} \right\rangle  - \left\langle {{\sigma _{22}}(t)} \right\rangle 
\end{equation}
which can be obtained either numerically or analytically with the use of Laplace transform, if the field amplitude is not pulsed but can be assumed constant.

We are particularly interested in the behavior of the ionization probability as a function of the intensity, for various detunings from the intermediate resonance. In order to provide results pertaining to a realistic model, we apply our theory to the $3s \to\to 4d \to Continuum$ process in Na, as described in the next section.

The above theoretical model, with minor modifications, reduces to the case of two-photon near-resonant ionization. The form of the density matrix equations is the same. Since the coupling between the initial and excited state is mediated by a single-photon transition, the Rabi frequency now is proportional to the field amplitude and not the intensity. For the same reason the relaxation constant of the off-diagonal matrix element is now given by ${\tilde \gamma _{12}} = {1 \over 2}\left[ {\gamma  + \Gamma \left( t \right) + {\gamma _L}} \right]$, where the laser bandwidth ${{\gamma_L}}$ is not multiplied by 2, as it was in the previous 3-photon case.

As discussed in the next section, the model is  applied to the realistic two-photon process $6s \to 7p \to Continuum$ process in Cs.

\section{\label{sec:level3}Results and Discussion}

In this section, we present quantitative results from the application of our theoretical model to the two realistic processes described above, namely the $3s \to\to 4d \to Continuum$ transition in Na and the $6s \to 7p \to Continuum$ in Cs. The atomic parameters \cite{ref18,ref22} obtained through quantum defect theory are: for the Na transition, ${\omega _{21}} = 4.2845eV$, $\Gamma \left[ {Hz} \right] = 9.4I\left[ {W/c{m^2}} \right]$, $\Omega \left[ {Hz} \right] = 1.46 \times {10^3}I\left[ {W/c{m^2}} \right]$, $S\left[ {Hz} \right] = 179I\left[ {W/c{m^2}} \right]$ while for the Cs transition, ${\omega _{21}} = 2.665eV$, $\Gamma \left[ {Hz} \right] = 11I\left[ {W/c{m^2}} \right]$, $\Omega \left[ {Hz} \right] = 0.75 \times {10^7}\sqrt {I\left[ {W/c{m^2}} \right]}$, where $I$ is the field intensity. Note that the Stark shift for an "1+1" photon process such as the $6s \to 7p \to Continuum$, as well as the spontaneous decay rates for both transitions in the range of intensities considered  are negligible and can be neglected. The laser bandwidth used in our calculations is ${\gamma _L} = 2 \times {10^{13}}Hz$. 
\begin{figure}[H]
	\centering
		\includegraphics[width=8cm]{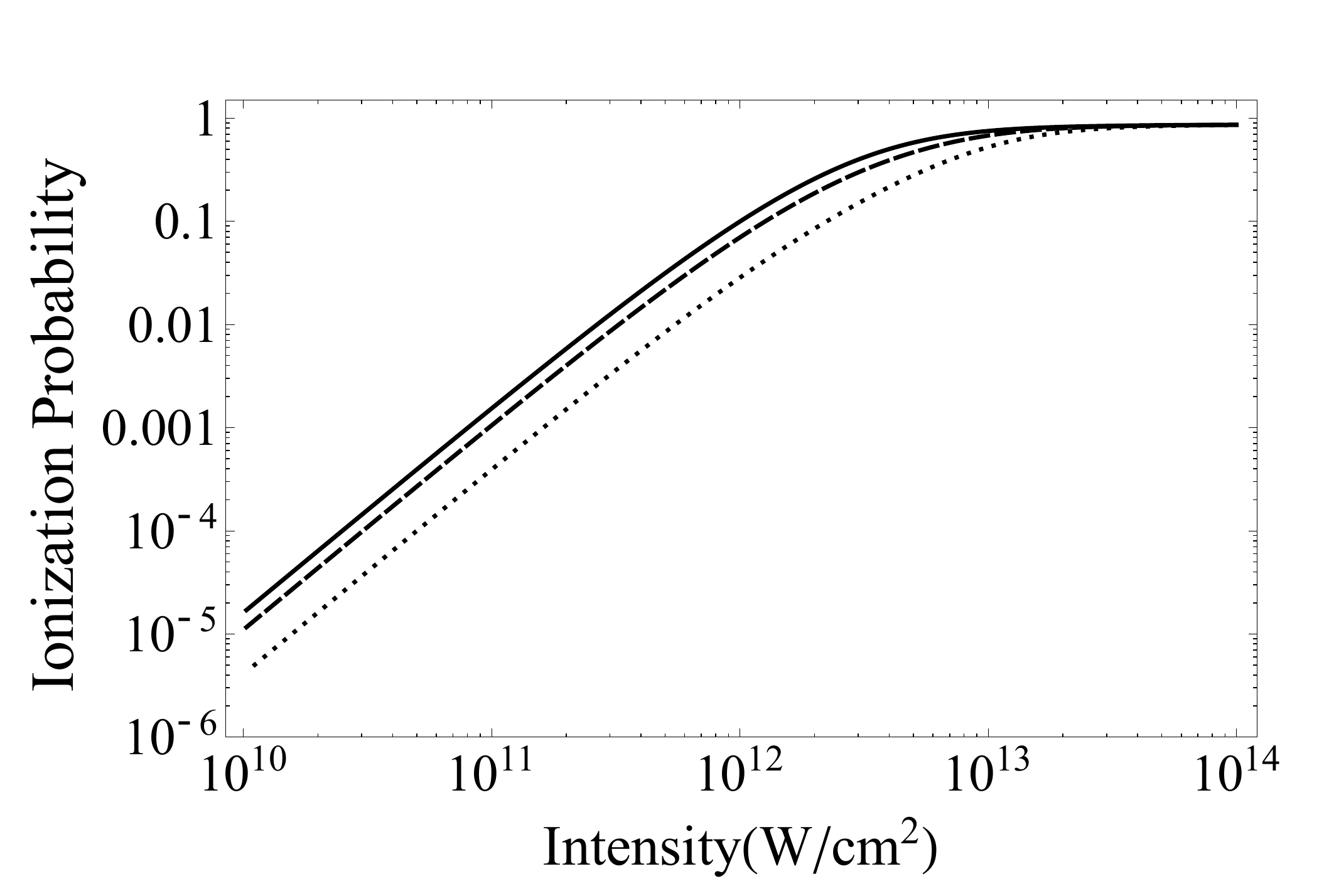}
	\caption[Cs100fs]{Ionization probability of Cs as a function of the intensity for various detunings from the intermediate resonance and $T=100fs$. Solid line: $\Delta /{\omega _2 } = 0.01$, dashed line: $\Delta /{\omega _2 } = 0.05$, dotted line: $\Delta /{\omega _2 } = 0.1$.}
\end{figure}

The illustrative results are summarized in figures 1 through 4. It should first be noted that the pulse duration is an additional parameter affecting the slope of  the ionization signal as a function of intensity. Obviously, for any peak intensity, if the pulse duration is sufficiently long, complete ionization will occur. In that limit, the slope will become zero, which will eventually be reached through a gradual decrease of the slope with increasing intensity, as shown in figures 1 and 2. A similar behaviour is shown in figures 3 and 4 with the difference that the slope becomes zero for intensities below the complete ionization regime. This indicates that  higher-order processes are more prone to depart from the non-resonant condition as we increase the intensity, compared to lower order ones.

\begin{figure}[H]
	\centering
		\includegraphics[width=8cm]{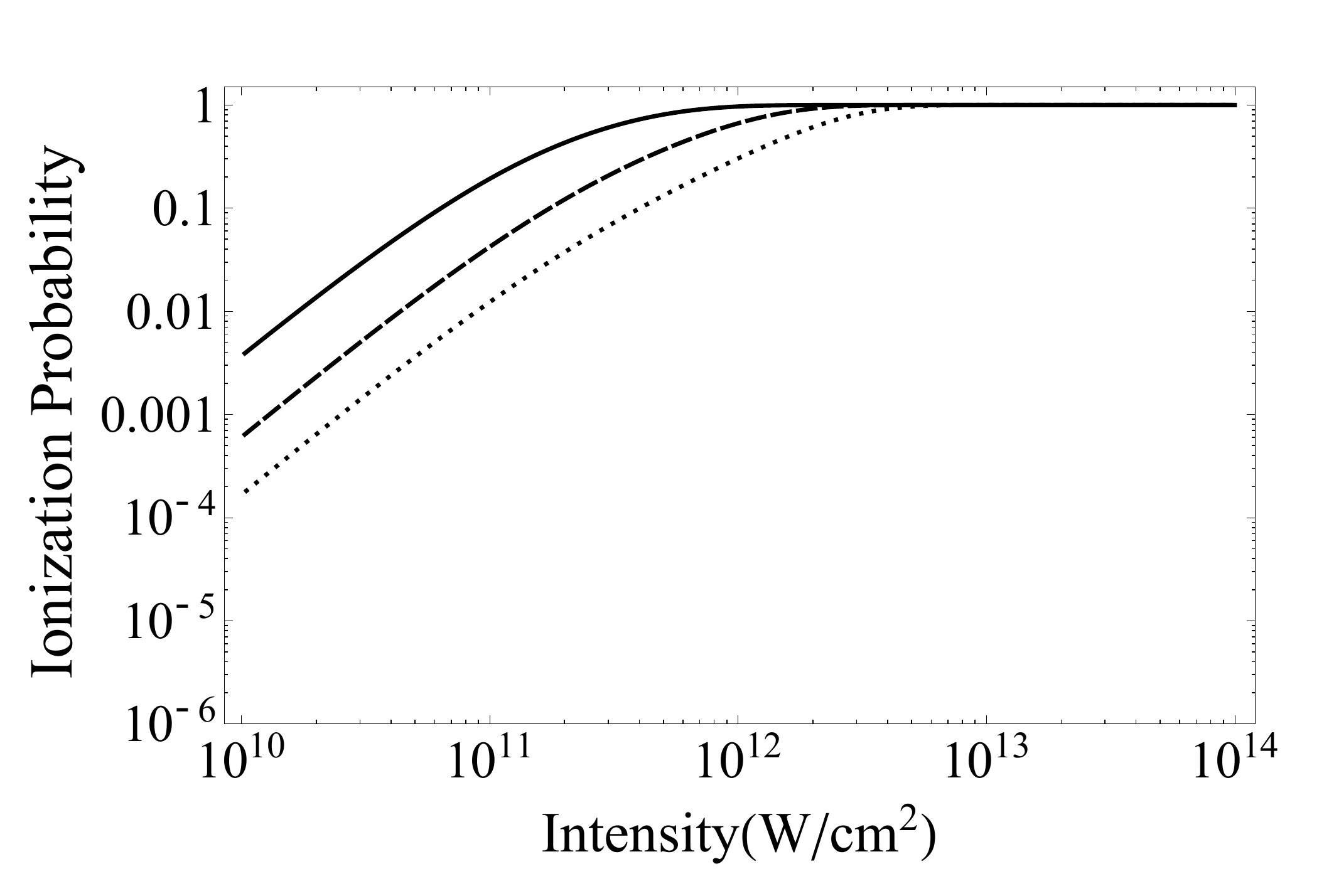}
	\caption[Cs1ps]{Ionization probability of Cs as a function of the intensity for various detunings from the intermediate resonance and $T=1ps$. Solid line: $\Delta /{\omega _2 } = 0.01$, dashed line: $\Delta /{\omega _2 } = 0.05$, dotted line: $\Delta /{\omega _2 } = 0.1$.}
\end{figure}

 Aside from this difference, the feature shared by all cases is that for larger detuning (dotted as compared to solid lines) from the intermediate resonance, the unperturbed slope reflecting the overall order of the process, persists for higher intensities. This behavior reflects the  onset of the effect of the increasing Rabi frequency, signaling the onset of departure from the notion of a non-resonant process. It is precisely the effect which in reference 1 was shown to herald the distortion of the photon statistics enhancement; although the process nominally is non-resonant. Therefore the message emerging from these results is that up to ${10^{10}}$ ${W/c{m^2}}$ or so, it can be assumed that the non-resonant behavior will persist. And that is the type of assessment that served as  the motivation for this paper.

It could be argued that the results may be of limited usefulness as they pertain to specific atomic transitions. 
However, they may not be as limited as they might seem at first sight. For one thing, parameters such as matrix elements entering in non-linear transitions do not differ by orders of magnitude. Cognizant of the specificity of our results, we do not claim exact limiting values of the intensities, but only a range of intensities. The validity
of that range is further underscored by comparison with the results of the recent data by Spasibko et al. \cite{ref2} who observed non-resonant slopes in harmonic generation up to order 4, for intensities in the above range. And that was in a completely different material. 

In view of the above estimates, in combination with the results in \cite{ref2}, and the dramatic progress in the potential for observation of non-linear processes induced by squeezed light, it appears that novel effects can be observed under intensities in the range of ${10^9}$ to 
${10^{10}}$ ${W/c{m^2}}$; as it is intensities in that range that are needed for the observation of such processes and are luckily becoming available even for squeezed radiation.

\begin{figure}[H]
	\centering
		\includegraphics[width=8cm]{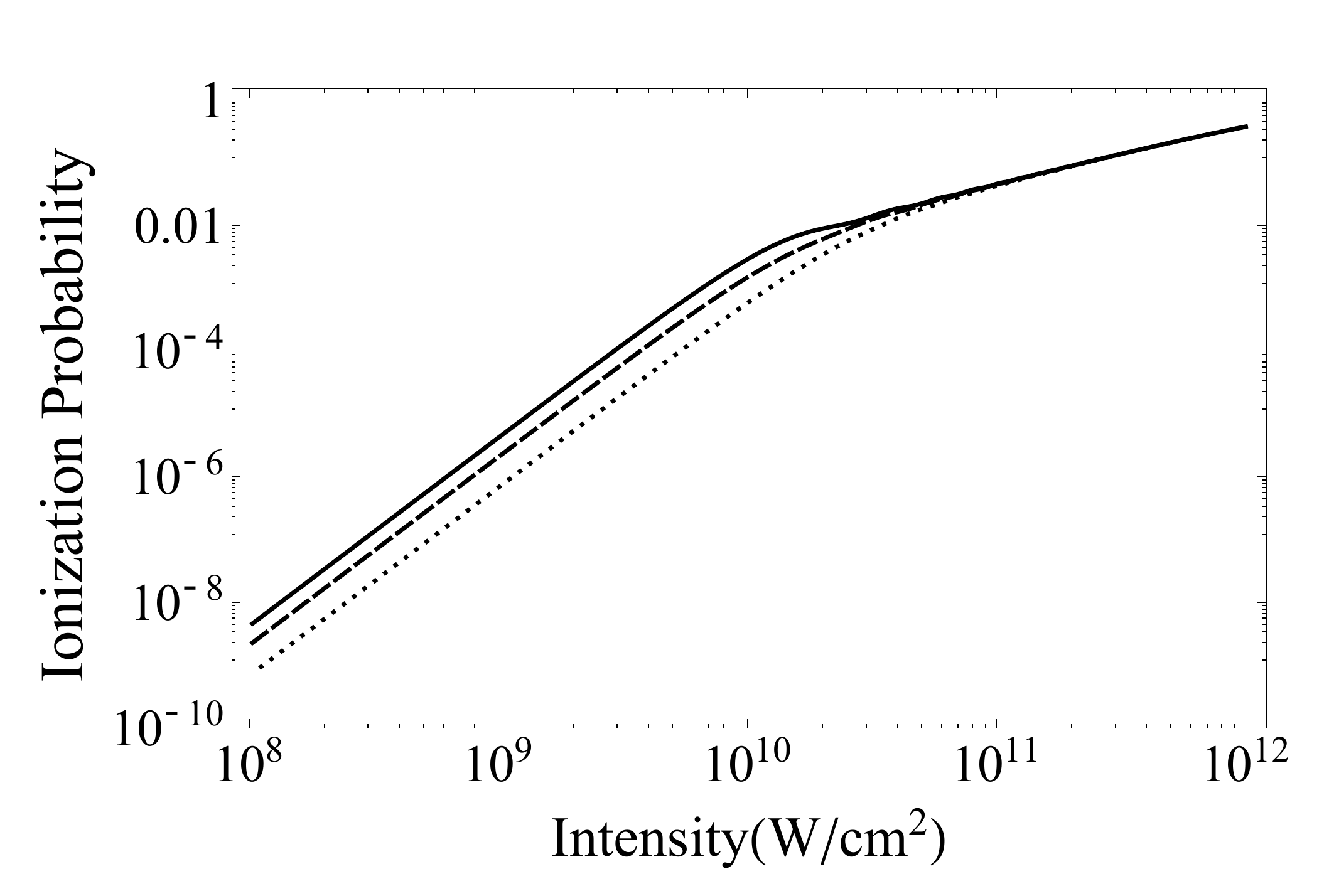}
	\caption[Na100fs]{Ionization probability of Na as a function of the intensity for various detunings from the intermediate resonance and $T=100fs$. Solid line: $\Delta /{\omega _2 } = 0.01$, dashed line: $\Delta /{\omega _2 } = 0.05$, dotted line: $\Delta /{\omega _2 } = 0.1$.}
\end{figure}

\begin{figure}[H]
	\centering
		\includegraphics[width=8cm]{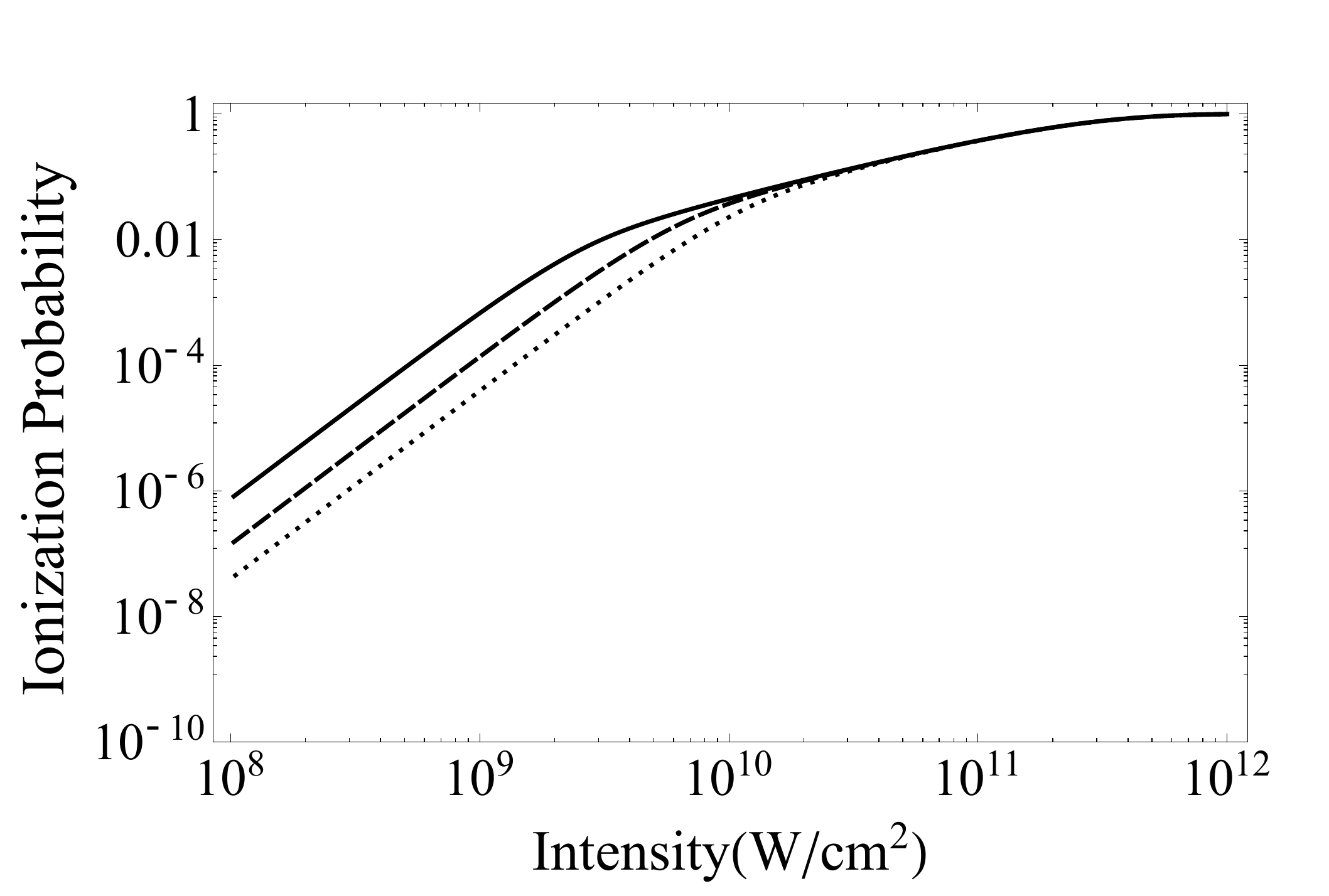}
	\caption[Na1ps]{Ionization probability of Na as a function of the intensity for various detunings from the intermediate resonance and $T=1ps$. Solid line: $\Delta /{\omega _2 } = 0.01$, dashed line: $\Delta /{\omega _2 } = 0.05$, dotted line: $\Delta /{\omega _2 } = 0.1$.}
\end{figure}

\end{document}